\documentclass[aps,pre]{revtex4}

 \usepackage{amsmath,amssymb,amsthm,mathrsfs,amsfonts,dsfont} 
\usepackage{dcolumn}
\usepackage{graphicx}
\usepackage{amsbsy}
\usepackage{bm}
\usepackage{algorithm}
\usepackage{algorithmicx}
\usepackage{algpseudocode}
\usepackage{physics}

\newcommand{\half}{\frac{1}{2}}
\newcommand{\lrbka}[1]{\left\langle #1 \right\rangle}
\newcommand{\G}{\mathcal{G}}
\def\i{\mathrm{i}}

\begin{document}

\title{An optimization-based equilibrium measure describes non-equilibrium steady state dynamics: application to edge of chaos}

\author{Junbin Qiu$^{1}$}
\author{Haiping Huang$^{1,2}$}
\email{huanghp7@mail.sysu.edu.cn}
\affiliation{$^{1}$PMI Lab, School of Physics,
Sun Yat-sen University, Guangzhou 510275, People's Republic of China}
\affiliation{$^{2}$Guangdong Provincial Key Laboratory of Magnetoelectric Physics and Devices, Sun Yat-sen University, Guangzhou 510275, People’s Republic of China}
\date{\today}

\begin{abstract}
Understanding neural dynamics is a central topic in machine learning, non-linear physics and neuroscience. However, the dynamics is non-linear, stochastic and particularly non-gradient, i.e., the driving force can not be written as gradient of a potential. These features make analytic studies very challenging. The common tool is the path integral approach or dynamical mean-field theory, but the drawback is that one has to solve the integro-differential or dynamical mean-field equations, which is computationally expensive and has no closed form solutions in general. From the aspect of associated Fokker-Planck equation, the steady state solution is generally unknown. Here, we treat searching for  the steady states as an optimization problem, and construct an approximate potential related to the speed of the dynamics, and find that searching for the ground state of this potential is equivalent to running an approximate stochastic gradient dynamics or Langevin dynamics. Only in the zero temperature limit, the distribution of the original steady states can be achieved. The resultant stationary state of the dynamics follows exactly the canonical Boltzmann measure. Within this framework, the quenched disorder intrinsic in the neural networks can be averaged out by applying the replica method, which leads naturally to order parameters for the non-equilibrium steady states. Our theory reproduces the well-known result of edge-of-chaos, and further the order parameters characterizing the continuous transition are derived, and the order parameters are explained as fluctuations and responses of the steady states. Our method thus opens the door to analytically study the steady state landscape of the deterministic or stochastic high dimensional dynamics.
\end{abstract}

\maketitle

\section{Introduction}
Non-equilibrium dynamics is a central topic of statistical physics~\cite{Risken-1996,Dyn-2018,RMP-2006}, from which general fluctuation-dissipation relations can be established~\cite{Kurchan-1998,Ohkuma-2007,Seifert-2012}, and
path integral approach was developed to derive the relationship between correlation and response function of the dynamics~\cite{MSR-1973,Jans-1976,Domin-1978}. One benchmark 
for testing or deriving theoretical hypotheses is the Langevin dynamics with conservative (expressed as gradient of a potential) or nonconservative forces, including Ising spin glass dynamics~\cite{Sommers-1987}, steady state thermodynamics~\cite{Sasa-2001}, and neural dynamics with partially symmetric couplings~\cite{Haim-1987}. These studies always resort to dynamical mean-field theory (DMFT)~\cite{Chow-2015,Roudi-2017}, where all possible trajectory paths are integrated to give a physical action, from which correlation and response functions are derived as an optimization of the action~\cite{Zou-2023}. In a general context, the dynamics would be very complicated, e.g., non-gradient and highly coupled, which makes the DMFT very complicated, and even it remains unknown how to capture the steady state properties of these dynamics.

It is commonly observed that the speed of dynamics determines whether the system enters a steady state and thus can be identified~\cite{NN-2013}. In this paper, we propose a new angle that only the low-speed or zero-speed region of the phase space is considered. In this sense, one can write down an energy cost function to optimize for satisfying this speed constraint. This optimization can be formulated under the stochastic gradient dynamics, where a temperature like parameter is introduced. This temperature plays the same role as in the canonical Boltzmann measure. In fact, in the zero temperature and stationary limit, the gradient dynamics under this approximate (or quasi) potential are self-consistent with the original general dynamics. Therefore, if the ground state of the energy cost is concentrated on, one can capture the steady state properties of the original phase space. This has an additional benefit that powerful equilibrium concepts such as Boltzmann measure, phase transitions, and order parameters can be leveraged~\cite{Huang-2022}.

We verify our framework in recurrent neural networks with asymmetric couplings, where the deterministic dynamics yields order-to-chaos transition~\cite{Chaos-1988,Brunel-2018}. We are able to handle the quenched disorder due to the coupling statistics with the help of replica method, a powerful trick in spin glass theory~\cite{Mezard-1987}. Furthermore, two types of order parameters are naturally derived; one describes the network activity, while the other describes the response property of the dynamic system. The former order parameter exhibits a continuous phase transition from null to positive value as the network parameters are tuned, while the latter one shows a peak at the exact location of transition, thereby offering rich physical information about the steady state of the complicated non-gradient dynamics.  Our framework also makes it possible to analytically study the quasi-potential landscape of a general dynamics, using large-deviation principle and landscape analysis in equilibrium statistical mechanics~\cite{LDP-2009,Huang-2022}, which is left for a future work~\cite{Wang-2024}.

Our quasi-potential analysis is not restricted to a simple type of dynamics, and can be generalized to treat competition dynamics of many species 
in ecosystems~\cite{RMP-1971,May-1972,Fyodorov-2016}, high-dimensional dynamics of random optimization~\cite{Fyodorov-2022,Vivo-2024}, high-dimensional gradient flow in machine learning~\cite{Max-2011}, and adaptive, hierarchical and noisy neural dynamics in neuroscience~\cite{ND-2014,Helias-2018,Helias-2020,Clark-2023,Jiang-2023}, as long as the dynamics speed can be identified and the steady state exists. The only limitation is that transient behavior before the steady state is reached could not be captured by our equilibrium approximation, and in that case, integro-differential equations are required to solve with expensive computational cost~\cite{Zou-2023}.  We next show how physical insights about a complicated non-gradient dynamics can be derived using our quasi-potential analysis.

\section{Model}
\begin{figure}
\centering
\includegraphics[width=0.95\textwidth]{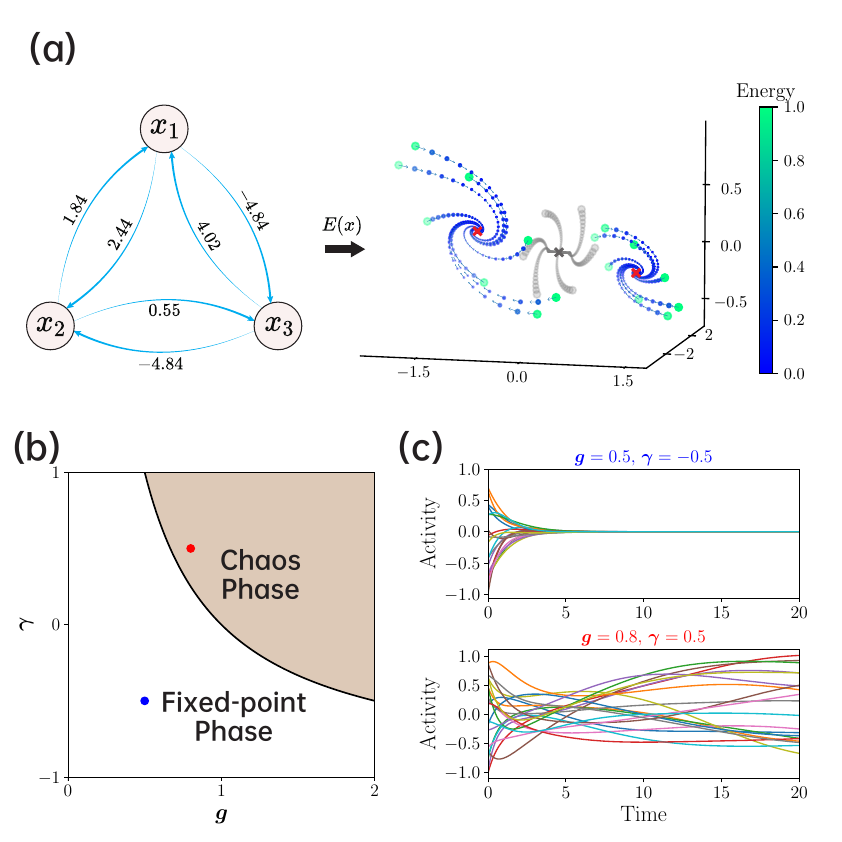}
\caption{Key idea of our quasi-potential method and phase diagram of the considered RNN model. 
(a) A three-neuron recurrent neural network is shown in the left panel.  Neurons are bidirectionally connected with the strength displayed near the arrows (synaptic directions).
Both connections in opposite directions are correlated once $\gamma \neq 0$.
Here, we simulate the RNN dynamics [Eq.~\eqref{eq: dynamicEquation}], whose phase space displays two fixed points ($g=1.2,\gamma=0.1$).
In the right panel, we show the trajectories flowing to these fixed points, where the color and size of the dots represent the energy of the corresponding position in the phase space,  and the arrows represent the direction and the $\ell_2$ norm of the velocity (the length of the arrow).
We also show an example of the trivial fixed-point for comparison ($g=0.8,\gamma=0.1$), where nearby trajectories are shown in the light grey color.
The cross symbols denote these corresponding fixed points.
(b) The phase diagram as a function of gain parameter $g$ and symmetric degree $\gamma$ is divided into two regions: null fixed-point and chaotic phases, separated by an analytic transition line $g(1+\gamma)=1$~\cite{Brunel-2018}. 
(c) We show the representative behavior of the dynamics [$\phi(x)$] in the two phases, corresponding to two colored points in the phase diagram of (b). }
\label{figure01}
\end{figure}

We consider a recurrent neural network (RNN) composed of $N$ interacting neurons, whose architecture is shown in Fig.~\ref{figure01} (a), illustrated as a simple example of $N=3$. 
The network activity is described by $N$ real-valued synaptic currents ($\mathbf{x}$). In the following, we use the bold symbol to indicate a vector or a matrix.
The connection strength $J_{i j}$ measures the asymmetric impact from neuron $j$ to neuron $i$.
Note that in our model, two neurons do not have to interact with each other equally, i.e. $J_{i j} \neq J_{j i}$. In addition, $J_{ii}=0$.  All neurons are connected to each other. The synaptic currents are transferred to firing rates through the current-to-rate transfer function $\phi(\cdot)$. Each neuron integrates neighboring activity via the incoming asymmetric synapses, and therefore the dynamics of the current is expressed as the following first-order non-linear differential equation for all $i$~\cite{Amari-1972,Chaos-1988}
\begin{equation}\label{eq: dynamicEquation}
    \dv{x_i}{t} = -x_i + \sum_{j = 1}^N J_{i j} \phi(x_j),
\end{equation}
where we consider unit time scale. Unless otherwise stated, $\phi=\tanh$. In this case, the transferred neural activity may become negative, while for biological reality, the non-negative function form should be chosen. However, our following analysis is still applicable to this type of non-negative transfer function.

In real neural circuits, $J_{ij}$ may not be identically independently distributed (i.i.d.). We therefore consider the correlated case whose statistics is specified below
\begin{align}
    \lrbka{\qty(J_{ij})^2} =& \frac{g^2}{N},\label{eq: variance1} \\
    \lrbka{J_{ij} J_{ji}} =& \frac{g^2}{N} \gamma,\label{eq: variance2}
\end{align}
where $g > 0$ is called the gain parameter, which controls the variance of connections, and $\gamma \in [-1, 1]$ controls the degree of connection symmetry. 
The special cases of $\gamma = 1$ and $\gamma = -1$ correspond to symmetric and anti-symmetric neural networks, respectively, while $\gamma = 0$ corresponds to fully asymmetric networks~\cite{Chaos-1988}. 

The above dynamics model exhibits an order-to-chaos transition~\cite{Brunel-2018}. The average (over the coupling realizations) number of equilibria explodes as the synaptic gain exceeds a critical value~\cite{PRL-2013}. The critical gain value can be derived from the rightmost boundary of the asymptotic eigenvalue spectrum of $\mathbf{J}$~\cite{Stein-1988}. When $g(1 + \gamma) < 1$, the network stays in the trivial fixed-point phase, where all trajectories flow to the zero-activity stable state. 
Once $g(1 + \gamma) > 1$, the network activity enters a chaotic phase, where an infinitesimal separation of two trajectories would be eventually amplified, and individual neurons exhibit slow to fast time scales as the network parameters go deep into the chaotic phase. 
We illustrate the phase diagram in Fig.~\ref{figure01} (b) and show the representative dynamic behaviors of these two phases in Fig.~\ref{figure01} (c).

Intuitively, the phase space is partitioned into different regions of different speeds ($\|\dv{\mathbf{x}}{t}\|^2$). We are interested in the zero-speed or low-speed regions~\cite{NN-2013}. 
Focusing on the speed landscape, we design the following approximate potential (quasi-potential) of the dynamics,
\begin{equation}\label{potential}
    E(\mathbf{x}) = \half \sum_i\qty(-x_i + \sum_{j = 1}^N J_{i j} \phi(x_j))^2 + \eta \norm*{\mathbf{x}}^2,
\end{equation}
where the first term is the kinetic energy, while the second term is a regularization term constraining the $\ell_2$ norm of the network activity.
$\eta$ plays the role of tuning the relative strength with respect to the kinetic energy. Here, we transform the studies of non-equilibrium steady states of non-gradient dynamics into an optimization of searching for the low-speed or zero-speed regions of the phase space, which corresponds to local or global minima of the above quasi-potential. This potential is equivalent to Hamiltonian in equilibrium statistical physics.

According to Eq.~\eqref{potential}, we can write down the following stochastic gradient dynamics (or Langevin dynamics)
\begin{equation}\label{LEQ}
\dv{\mathbf{x}}{t}=-\nabla_{\mathbf{x}}E(\mathbf{x})+\sqrt{2T}\boldsymbol{\epsilon},
\end{equation}
where $\boldsymbol{\epsilon}(t)$ is a time-dependent white noise, whose statistics is given by $\lrbka{\epsilon_i(t)}=0$, $\lrbka{\epsilon_i(t)\epsilon_j(t')}=\delta_{ij}\delta(t-t')$. The temperature $T$ is used to tune the energy level, playing the same role as in the equilibrium Boltzmann measure. We can write the gradient (force) in Eq.~\eqref{LEQ} in the component wise, i.e., $-\frac{\partial E(\mathbf{x})}{\partial x_i}\equiv F_i$ where
\begin{equation}
F_i\equiv-x_i+h_i-\phi'(x_i)\sum_{j:j\neq i}J_{ji}(h_j-x_j),
\end{equation}
where $h_i\equiv\sum_{j:j\neq i}J_{ij}\phi(x_j)$, and the last term explains the feedback effects due to the neuron $i$ whose activity yields impacts on its neighbors. From this expansion, we can see that only in the stationary limit ($T\to 0$ as well), the above Langevin dynamics closely resembles the original dynamics (the same magnitude order of small speeds). In this sense, the energy defined in Eq.~\eqref{potential} is not a Lyapunov function and even the steady-state potential whose analytic form for the current out-of-equilibrium dynamics with asymmetric couplings is unknown~\cite{AP-2004}, although the driving force is argued to be decomposed into the gradient of a potential related to the steady state distribution and a divergent-free curl flux~\cite{WJ-2013}.
The decomposition takes into account the probability conservation of neural state evolution. In contrast, we start from an optimization angle that concentrates on the low-speed region of the phase space, and formulate the steady state behavior as the canonical ensemble of stationary fixed points. 

It is well-known that the steady state of the above Langevin dynamics has an equilibrium Boltzmann measure~\cite{Risken-1996},
\begin{equation}\label{BM}
P(\mathbf{x})=\frac{1}{Z}e^{-\beta E(\mathbf{x})},
\end{equation}
where the partition function $Z$ depends on particular realizations of $\mathbf{J}$, and $\beta=1/T$. An illustration of $E(\mathbf{x})$ is given in Fig.~\ref{figure01} (a). Our main focus is to compute the free energy of this canonical ensemble, and derive the steady state properties of the dynamics model. The theoretical predictions can be confirmed by simulating the discretized Langevin dynamics, assuming the small discretization step as learning rate, akin to a Monte-Carlo simulation. We remark that this equilibrium analysis reproduces the continuous nature of the dynamics transition, more insights and possibility of exploring the speed landscape would be discussed in the remaining sections.

\section{Replica method}
To derive the quenched disorder average of the free energy, we apply the replica method, which is popular in studying theory of spin glasses~\cite{Mezard-1987} and statistical mechanics of neural networks~\cite{Huang-2022}. We first write down the following replicated partition function
\begin{equation}\label{eq: partitionFunction}
\begin{aligned}
    Z^n &= \int d \mathbf{x} \exp\left[ -\beta \pqty{\half \sum_{ia} \pqty{-x_i^a + \sum_j J_{ij} \phi(x_j^a)}^2 + \eta\sum_a ||\mathbf{x}^a||^2}\right],\\
    & =\int d \mathbf{x} D \mathbf{\hat{x}} \exp \bqty{\i \sqrt{\beta} \sum_{ia} \hat{x}_i^a \pqty{-x_i^a + \sum_j J_{ij} \phi(x_j^a)} - \beta \eta\sum_a \norm*{\mathbf{x}^a}^2},
    \end{aligned}
\end{equation}
where to arrive the second equality, the Hubbard–Stratonovich transformation is used~\cite{Huang-2022}, and $i,a$ denote the neuron and replica index, respectively. The neuron index runs from $1$ to $N$, while the replica index runs from $1$ to $n$. We write a shorthand $D\mathbf{\hat{x}}\equiv\prod_{i = 1}^N \prod_{a = 1}^n D x_i^a$ where   $D \hat{x} \equiv e^{-\half \hat{x}^2} d \hat{x}/ \sqrt{2 \pi}$ being the Gaussian measure, and we call $\hat x$ as a response field~\cite{Zou-2023}. We also denote $d \mathbf{x}$ as a shorthand for $\prod_{i = 1}^N \prod_{a = 1}^n d x_i^a$, where $x_i^a\in\mathbb{R}$. Therefore, $-\beta f=\lim_{n\to0}\frac{\ln\lrbka{Z^n}}{Nn}$~\cite{Mezard-1987}, where $\lrbka{\cdot}$ denotes the disorder average over $\mathbf{J}$. The neural dynamics is much faster than the coupling dynamics, in which we assume a stationary limit of fixed $\mathbf{J}$, and thus the quenched disorder average is required. But if both dynamics evolve in similar time scales, an enlarged state space may be considered, and we leave this issue to future works.

In the course of replica analysis (see details in Appendix~\ref{app-rep}), two order parameters are naturally introduced as follows,
\begin{align}
    Q^{ab} =& \frac{1}{N} \sum_i \phi(x_i^a) \phi(x_i^b), \\
    R^{ab} =& \frac{1}{N} \sum_i \hat{x}_i^a \phi(x_i^b),
\end{align}
where $\hat{x}_i^a$ acts as a response field, whose name is inspired by the analysis using DMFT~\cite{Zou-2023} and explained below. Briefly, this response field emerges due to the linearization of the quadratic terms in the Hamiltonian [see Eq.~\eqref{potential} and Eq.~\eqref{eq: partitionFunction}].
As the first level of approximation, we assume the replica symmetric ans\"atz to order parameters $Q^{ab} = q \delta_{ab} + Q(1 - \delta_{ab})$ and $R^{ab} = r \delta_{ab} + R(1 - \delta_{ab})$, whose physical meaning is explained in more details in Appendix~\ref{app-rep}.
Under this approximation, the free energy is given by
\begin{equation}\label{fT}
\begin{aligned}
    -\beta f =& \frac{1}{2} Q \hat{Q} - q \hat{q} +  R \hat{R} - r \hat{r} - \ln \sigma + \frac{1}{2} \beta g^2 \gamma \qty(r^2 - R^2) \\
    & + \int \pqty{D u D v} \ln I,
\end{aligned}
\end{equation}
where $\sigma \equiv \sqrt{1 + g^2 \beta (q-Q)}$, and $\hat{Q}$, $\hat{q}$, $\hat{R}$ and $\hat{r}$ are conjugated order parameters. 
 The integral $I \equiv \int d x e^{\mathcal{H}(x)}$, where
\begin{equation}
\begin{aligned}
    \mathcal{H} (x) \equiv& - \beta \eta x^2 + \frac{1}{2} \qty(2 \hat{q} - \hat{Q}) \phi^2(x) \\
    & + \qty(\sqrt{\frac{g^2 \beta Q \hat{Q} - \hat{R}^2}{g^2 \beta Q}} u + \frac{\hat{R}}{g \sqrt{\beta Q}} v) \phi(x) \\
    & - \frac{1}{2 \sigma^2} \qty(g \sqrt{\beta Q} v + (\hat{r} - \hat{R}) \phi(x) - \sqrt{\beta} x)^2.
\end{aligned}
\end{equation}

In the large $N$ limit, we can use the saddle-point approximation (also called Laplace method) to derive the closed equations the order parameters 
must obey. Therefore, by setting the derivatives of the replica free energy with respect to the order parameters zero, we obtain the following compact self-consistent equations (see Appendix~\ref{app-sde} for details)
\begin{subequations}\label{sde-rnn}
\begin{align}
    q =& \qty[\lrbka{\phi^2}],\\
    Q =& \qty[\lrbka{\phi}^2],\\
    \hat{q} =& -\frac{g k}{2} + \frac{g^2 k^2 Q}{2} + \frac{g k}{2 \sigma^2} \qty(\hat{r} -\hat{R}) f(1, 1, -2) \qty[\lrbka{\phi^2}] \\
    + &\frac{g k}{\sigma^2} \qty(\hat{r} -\hat{R}) f(0, -1, 1) \qty[\lrbka{\phi}^2] + \frac{k^2}{2} \qty(1 - 2 g k Q) \qty[\lrbka{x^2}] \\
    + &\frac{g k \sqrt{\beta}}{\sigma^2} f (0, 1, -2) \qty[\lrbka{x}\lrbka{\phi}] + \frac{g k \sqrt{\beta}}{\sigma^2} f(-1, 0, 2) \qty[\lrbka{x \phi}] \\
    + & g k^3 Q \qty[\lrbka{x}^2],\\
    \hat{Q} =& g^2 k^2 Q + \frac{2 g k}{\sigma^2} \qty(\hat{r} -\hat{R}) f(0, 1, -1) \qty[\lrbka{\phi^2}] \\
    + &\frac{g k}{\sigma^2} \qty(\hat{r} -\hat{R}) f(1, -3, 2) \qty[\lrbka{\phi}^2] - 2 g k^3 Q \qty[\lrbka{x^2}] \\
    - &\frac{2 g k \sqrt{\beta}}{\sigma^2} f(1, -2, 2) \qty[\lrbka{x}\lrbka{\phi}] + \frac{2 g k \sqrt{\beta}}{\sigma^2} f(0, -1, 2) \qty[\lrbka{x \phi}] \\
    + &k^2 \qty(1 + 2 g k Q) \qty[\lrbka{x}^2],\\
    r =& -\frac{1}{\sigma^2} f(1, 0, -1) \qty[\lrbka{\phi^2}] + \frac{1}{\sigma^2} f(0, 1, -1) \qty[\lrbka{\phi}^2] \\
    + &\frac{\sqrt{\beta}}{\sigma^2} \qty(1 - g k Q) \qty[\lrbka{\phi x}] + \frac{\sqrt{\beta}}{\sigma^2} g k Q \qty[\lrbka{\phi}\lrbka{x}],\\
    R =& -\frac{1}{\sigma^2} f(0, 1, -1) \qty[\lrbka{\phi^2}] - \frac{1}{\sigma^2} f(1, -2, 1) \qty[\lrbka{\phi}^2] \\
    - &\frac{\sqrt{\beta}}{\sigma^2} g k Q \qty[\lrbka{x \phi}] + \frac{\sqrt{\beta}}{\sigma^2} \qty(1 + g k Q) \qty[\lrbka{x}\lrbka{\phi}],\\
    \hat{r} =& \beta g^2 \gamma r,\\
    \hat{R} =& \beta g^2 \gamma R,
\end{align}
\end{subequations}
where $k \equiv \frac{g \beta}{\sigma^2}$ and $f(a, b, c) \equiv a \hat{r} + b \hat{R} + c g k Q \qty(\hat{r} - \hat{R})$. In addition, two kinds of averages are specified. One is the disorder average $\qty[\cdot] \equiv \int Du Dv \cdot$,  and the other is the thermal average $\lrbka{ \cdot } \equiv \frac{\int dx e^{\mathcal{H}(x)} \cdot}{\int dx e^{\mathcal{H}(x)}}$. $\mathcal{H}(x)$ plays the role of effective Hamiltonian of our model. We emphasize that this thermal average is only defined in this saddle-point equation, or other quantities that can be expressed as a function of order parameters; otherwise $\langle\cdot\rangle$ should be referred to the average over the coupling statistics. When $\gamma=0$ (i.i.d. scenario), $\hat{r},\hat{R}$ and $f(a,b,c)$ vanish, the above saddle point equation can be greatly simplified (see Appendix~\ref{app-0}).

We finally remark that the above two kinds of order parameters arising from the quenched disorder average have clear physical meanings. First, $q-Q$ characterizes actually the activity fluctuation in the neural population, which can be directly measured in neural circuits. Second, $R^{ab}$ can be interpreted as the response function. More precisely, the steady state receives a perturbation of synaptic currents (e.g, denoted by $I_a$, where $a$ can be assumed as a state index~\cite{Mezard-1987}), and the linear response $\frac{\partial{\langle\phi_b\rangle}}{\partial I_a}$ characterizes how the system reacts to the small perturbation. $\phi_b$ is the mean-field activity of the state $b$. Note that the average $\langle\cdot\rangle$ is defined under Eq.~\eqref{eq: partitionFunction}, which leads exactly to $R^{ab}=\langle\hat{x}^a\phi_b\rangle$. Therefore we explain $R^{ab}$ as a linear response in physics. We shall support this picture by theoretical arguments and numerical simulations in the next section.

\section{Results and discussion}
\subsection{Equilibrium replica analysis reveals the continuous nature of the chaos transition}
The zero temperature limit selects the ground state of the Boltzmann measure [Eq.\eqref{BM}], which captures the statistics of the steady state solutions of  the original non-gradient dynamics [Eq.~\eqref{eq: dynamicEquation}]. An explicit form of Lyapunov function that is monotonically decreasing over time can only be identified provided that the neural coupling is symmetric, and only in this special case, the Lyapunov function reduces to an Ising-model like Hamiltonian, as encountered in standard Hopfield networks~\cite{Cool-2001}. We thus remark that the quasi-potential defined in Eq.~\eqref{potential} is not the physical potential of the original dynamics, yet capturing only the steady state properties of the dynamics, e.g., the low- (or zero-) speed landscape. In this sense, not all characteristics of a non-gradient dynamics are included in our approximate potential description, e.g., the transient behavior or the temporal evolution of autocorrelation even in the time translation invariant regime is not captured. However, the statistics of fixed-point solutions can be well captured by the quasi-potential approach.

\begin{figure}[ht]
    \centering
    \includegraphics[width=0.95\textwidth]{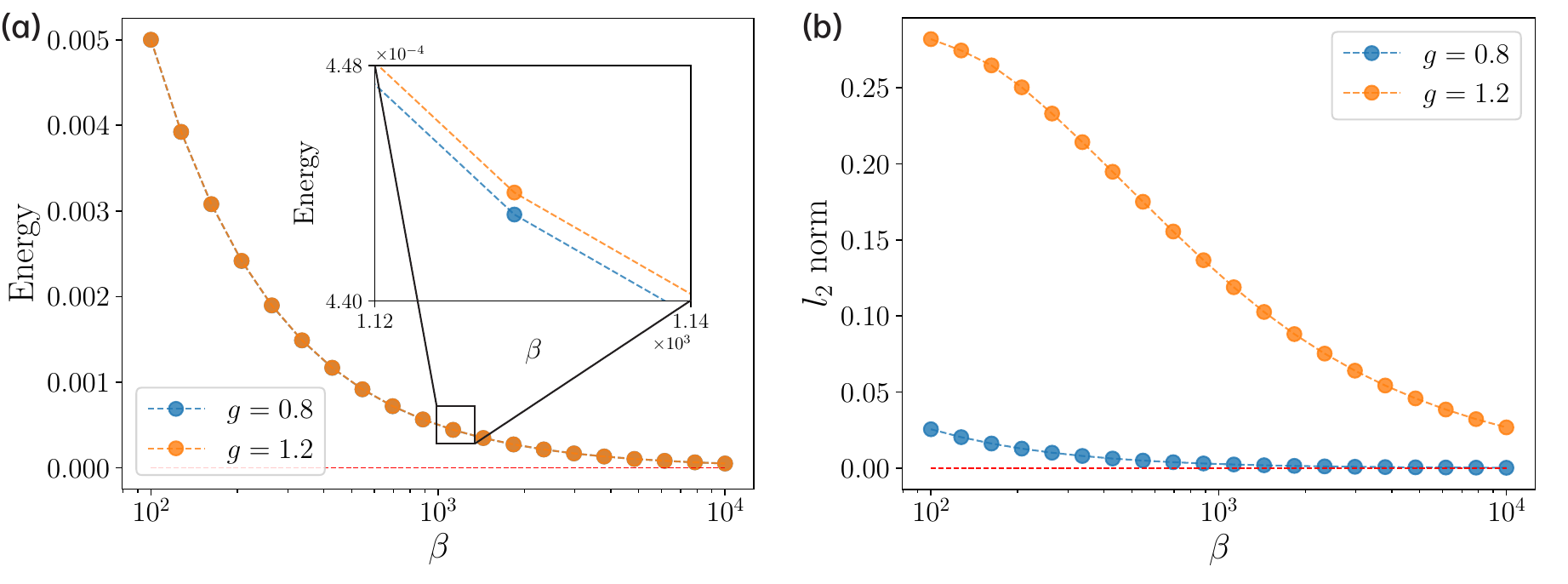}
    \caption{The average energy (a) and $\ell_2$ norm (b) of network activity versus the inverse temperature $\beta$ in the i.i.d. scenario $\gamma=0$ with $\eta=0$. 
    The orange and the blue curves represent the setting $g=0.8$ and $g=1.2$, respectively.
    The red dashed line denotes the ground state energy. Two lines of $g=0.8$ and $g=1.2$ are nearly indistinguishable. }
    \label{energy}
\end{figure}

The average energy is an indicator of whether the RNN dynamics is trapped in fixed points. In other words, steady states are achievable provided that the average energy can be optimized to the ground state zero value. Without loss of generality, we consider the independent setting $\gamma=0$ and set $\eta = 0$. As shown in Fig.~\ref{energy}, with increasing value of $\beta$,
the energy decreases to $0$, implying that the ground state can be achieved in a very large $\beta$. 
When $\beta = 10000$, the energy becomes less than $10^{-4}$, suggesting that it is reasonable to set $\beta=10000$ for the zero temperature limit considered in our numerical solutions of the self-consistent equation [see Eq.~\eqref{sde-rnn}] and experiments verifying our theoretical results. We also show the $\ell_2$ norm of the network activity in Fig.~\ref{energy}. The norm displays different magnitudes before and after the onset of chaos.

\begin{figure*}[t]
    \centering
    \includegraphics[width=1.0\textwidth]{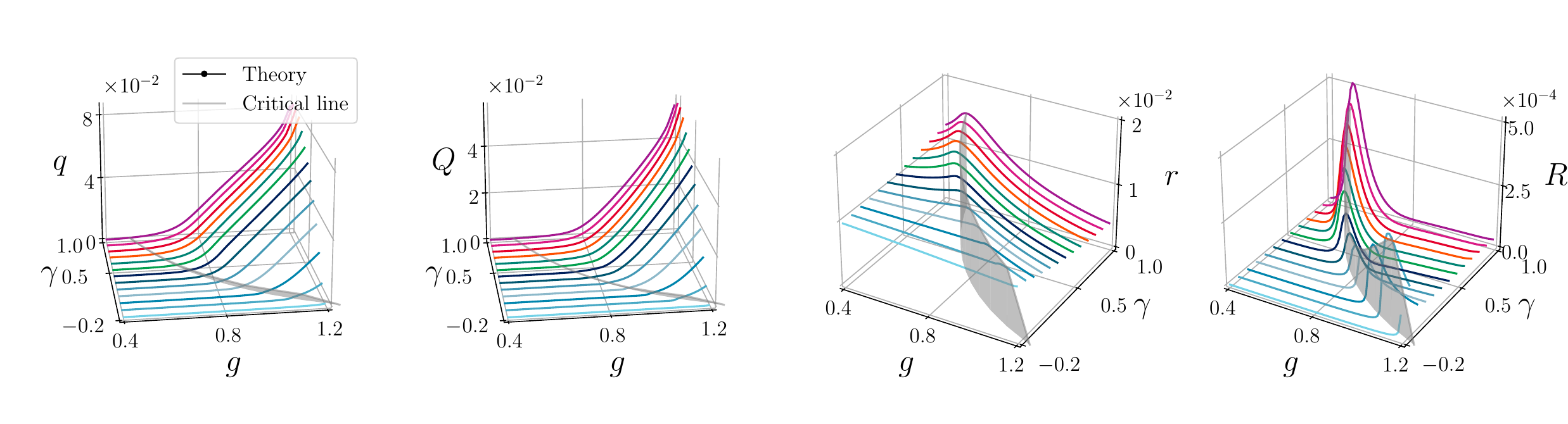}
    \caption{Order parameter profile as a function of gain parameters $g$ and symmetric degree $\gamma$. $g$ is varied from $0.4$ to $1.2$, and $\gamma$ is varied from $-0.2$ to $1.0$ with an interval of $0.1$. The colors of curves correspond to the different $\gamma$ values. The grey vertical plane specifies the location of the critical line $g \qty(\gamma + 1) = 1$~\cite{Brunel-2018}.}
    \label{fig-PD}
\end{figure*}

With a finite temperature, the low-speed solutions of the original dynamics could also be studied. For the remaining analysis, we focus on a very large $\beta$ (e.g., $\beta=10^4$) and set $\eta=0$ unless otherwise stated. Varying the control parameters $g$ and $\gamma$, we plot the profiles of all order parameters ($q,Q,r,R$) in Fig.~\ref{fig-PD}.

Our replica analysis reveals exactly the order-to-chaos transition~\cite{Brunel-2018}. When $g<\frac{1}{1+\gamma}$, the zero fixed-point uniquely dominates the phase space, resulting in vanishing $q$ and $Q$, as expected from the physical definition of the order parameters. Once $g>\frac{1}{1+\gamma}$, $q$ and $Q$ increases rapidly but continually from zero, indicating the emergence of self-sustained collective activity. This coincides with the known picture of exponentially growing topological complexity~\cite{PRL-2013} and the positive maximal Lyapunov exponent~\cite{Chaos-1988,Helias-2020,Huang-2022}. It would be interesting to count the stationary points of our potential, rather than using the hard-to-compute Kac-Rice formula~\cite{PRL-2013}, which we left for a future thorough work~\cite{Wang-2024}. Our method thus provides a relatively simple way to access the dynamics landscape in the steady state limit. We remark that the steady state distribution of the network activity for the original non-gradient dynamics does not have an analytic form to date.

\subsection{Response order parameter for non-equilibrium steady states}

Let us now look at the associated order parameters involving the response field $\hat{x}$, which emerges due to the linearization of the quadratic terms in the Hamiltonian (see details in Appendix~\ref{app-rep}). This response field is similar to that used to enforce the dynamics equation in a field-theoretic formula of the stochastic dynamics~\cite{Zou-2023}. But in our current setting, the response field is time-independent, as only steady state properties are focused on in this work. Therefore, we argue that $R$ and $r$ bear the same physical meaning with the steady state integrated response function in the DMFT language~\cite{Zou-2023}. Let us explain in detail. Based on the derivation of DMFT in our recent work~\cite{Zou-2023}, 
we have the one-dimensional mean-field dynamics equation:
\begin{equation}\label{dmfteq}
    \dv{x(t)}{t} = - x(t) + \gamma g^2 \int_0^\infty R(t, t') \phi(t') d t' + \omega(t),
\end{equation}
where $\expval{\omega(t) \omega(t')} = g^2 C(t, t')$, and $C(t,t')=\langle\phi(t)\phi(t')\rangle$ where the average is done in the dynamical mean-field sense.

In the steady state, we have the property of time translation invariance $R(t,t')=R(t-t')$, and thus we can simplify the second term in the right hand side of Eq.~\eqref{dmfteq} as follows,
\begin{equation}
    \int_0^\infty R(t, t') \phi(t') d t' = \int_0^\infty R(u) \phi(t - u) d u \to R_{\text{int}} \phi^*,
\end{equation}
where $R_{\text{int}} = \int_0^\infty R(u) d u$ is the time integral of $R(t)$, and $\phi^*$ denotes the steady state value.
Finally, we will obtain the following steady state solution,
\begin{equation}\label{eq: dmft-mfe}
    x^* = \omega^* + w \phi^*,
\end{equation}
where $w \equiv g^2 \gamma R_{\text{int}}$, and $\omega^*$ denotes a noise sampled from $\mathcal{N} \qty(0, g^2 C)$. For each noise sample, $x^*$ and $\phi^* \equiv \phi(x^*)$ denote the optimal values satisfying Eq.~\eqref{eq: dmft-mfe}. Therefore, correlation and response functions are self-consistently given by
\begin{equation}\label{creq}
\begin{aligned}
    C =& \expval{\qty(\phi^*)^2}_{\omega},\\
    R_{\text{int}} =& \expval{\phi'\qty(\omega^* + w \phi^*) \qty(1 + w R_{\text{int}})}_{\omega},
\end{aligned}
\end{equation}
where the average is done by drawing noise samples from $\mathcal{N} \qty(0, g^2 C)$.

\begin{figure}[htbp]
    \centering
    \includegraphics[width=0.9\textwidth]{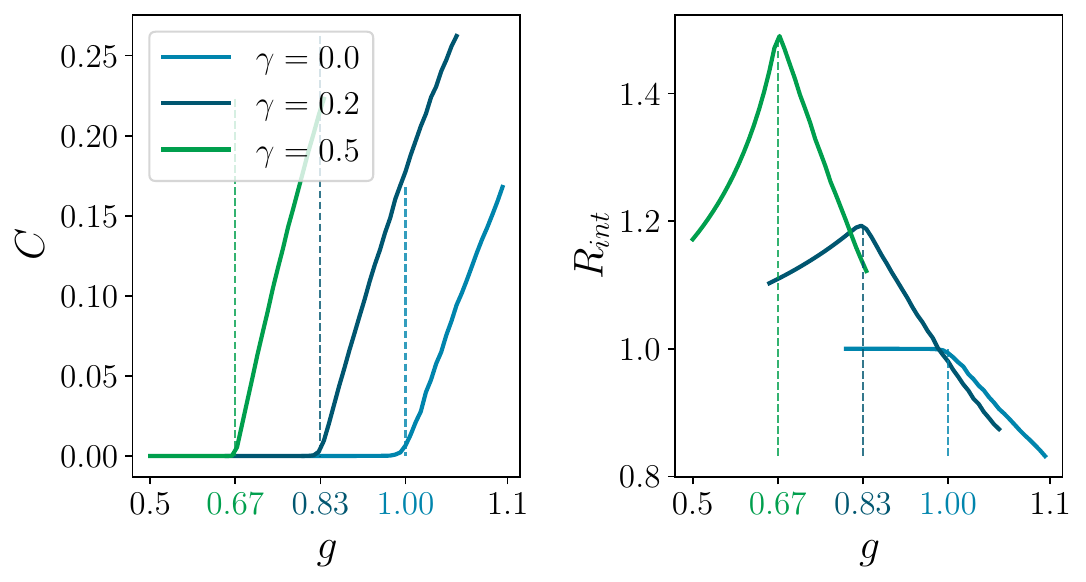}
    \caption{The behavior of correlation function $C$ and response function $R_{\text{int}}$ derived from the DMFT method~\cite{Zou-2023}. We choose $\gamma = 0.0, 0.2, 0.5$, and $g$ varies from $0.5$ to $1.1$ in the simulations. The vertical dashed lines denote the critical points for each $\gamma$ value.}
    \label{fig:dmft}
\end{figure}

A numerical iteration of Eq.~\eqref{creq} gives us the fixed-point of $(C,R_{\rm int}$), which is plotted in Fig.~\ref{fig:dmft}. This corresponds exactly to the behavior of $q$ (or $Q$), and even the behavior of the response order parameter derived using our quasi-potential method. We remark that our method takes an optimization angle, rather than the path integral angle. Therefore, it is reasonable to define and derive the order parameters in equilibrium ensemble theory (such as Boltzmann canonical ensemble here), and conclude the order-to-chaos transition is of the continuous type. Most interestingly, we can relate the replica overlap matrix to the observable neural activity, e.g., $q-Q$ measures the activity fluctuation, and $r-R$ measures the response of the system.

 In fact, $r$ and $R$ display a peak at the transition point. This peak is an indicator of continuous dynamics transition, and may be related to the divergence of the relaxation time-scale of the auto-correlation function~\cite{Chaos-1988,Brunel-2018}. Most interestingly, the decay behavior of the response function in both sides of the transition looks very different, and in the chaotic side, the response decays more slowly. An earlier work revealed that the memory lifetime of the network under a linear decoder decreases more slowly in the chaotic side than in the non-chaotic side~\cite{PRE-2011}. This shows that, the response order parameters revealed in our current work have a deep connection to other performance metrics of random neural networks, and they may share the same mathematical foundation. We thus conclude that the dynamics at the edge of chaos is more responsive than at other locations in the phase diagram. The edge of chaos has thus great computational benefits, even a bit far away from the transition point to the chaotic regime~\cite{EoC-1990,EoC-2004,Huang-2022}. Surprisingly, this is also consistent with the experimental data of cortical electrodynamics~\cite{PNAS-2022}, which claimed the information richness at the edge of chaos, and the conscious brain states are observed at the edge of chaos but \textit{not deep} in the chaotic regime (see Figure 2 in the recent work~\cite{PNAS-2022}). We conclude that this peak of the response function is a meaningful byproduct of our theory, and moreover an indicator of the edge of chaos. Taking a step further, it is a necessary condition for the emergence of consciousness.

A careful inspection of the saddle-point equation [see Eq.~\eqref{sde-rnn}] leads to the following concise relationship between $q-Q$ (fluctuation) and $r-R$ (response):
\begin{equation}
r-R=\frac{\sqrt{\beta}\sigma_{x\phi}}{1+\beta g^2(q-Q)(1+\gamma)},
\end{equation}
where $\sigma_{x\phi}=[\langle x\phi\rangle]-[\langle x\rangle\langle\phi\rangle]$, characterizing the current-activity correlation. Interestingly, the current-activity correlation behaves as an approximately linear function of the activity variance. The relation between the response and fluctuation is quite non-linear, and at some location, there exhibits a peak that indicates the transition. It is thus very interesting in future works to derive a non-equilibrium linear response theory.

\begin{figure}
    \centering
    \includegraphics[width=1.0\textwidth]{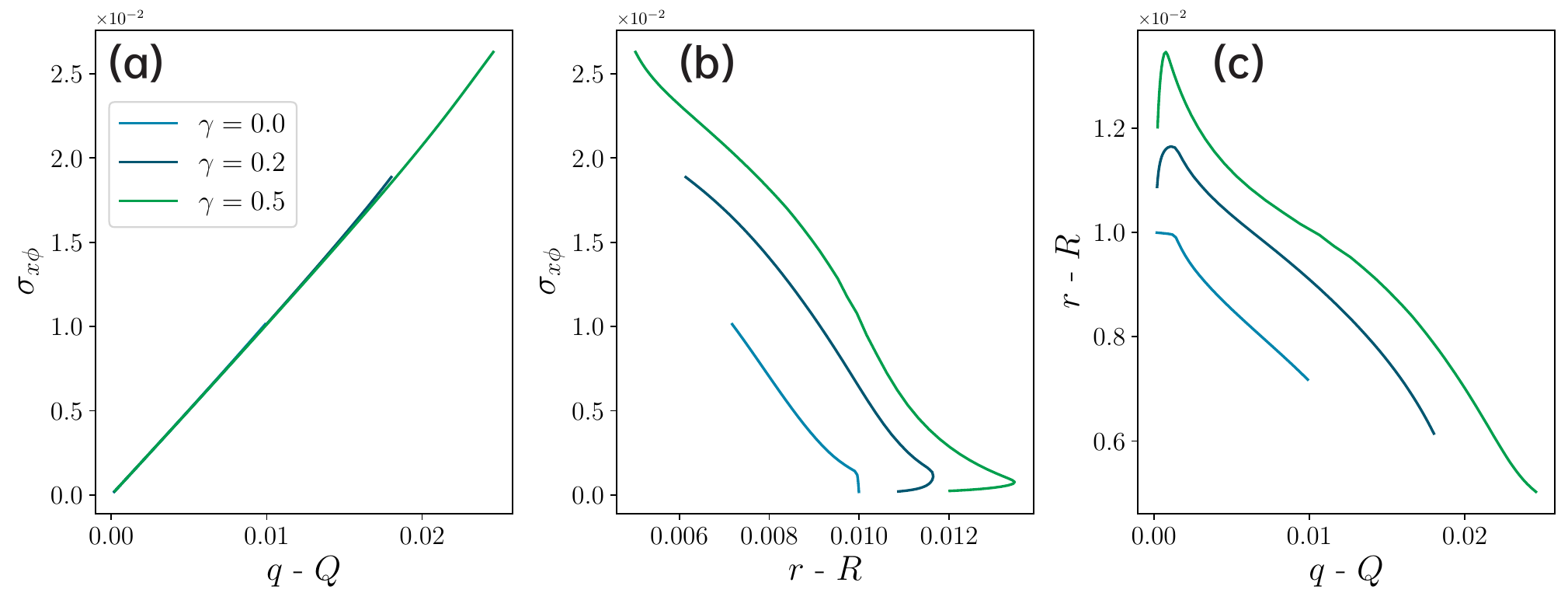}
    \caption{Fluctuation-response relationship for non-equilibrium steady states. The plotted order parameters follow the saddle-point equation derived in the main text [see Eq.~\eqref{sde-rnn}]. (a) $\sigma_{x\phi}$ vs $q-Q$. (b) $\sigma_{x\phi}$ vs $r-R$. (c)  $r-R$ vs $q-Q$.
   } 
    \label{FRR}
\end{figure}

\subsection{Scaling behavior of order parameters in the zero temperature limit}
In the zero temperature limit, we have to consider the following scaling behavior guaranteeing a finite value of free energy when $\beta\to\infty$ [see Eq.~\eqref{fT}].
\begin{equation}
\begin{aligned}
    q - Q&\to\frac{\chi}{\beta}, \\
    \hat{Q} &\to \beta^2 \hat{Q}, \\
    2 \hat{q} - \hat{Q}&\to-2\beta \hat{\chi},\\
    r&\to\sqrt{\beta}\tilde{r},\\
    R-r&\to\frac{\xi}{\sqrt{\beta}},\\
    \hat{R}&\to\beta^{\frac{3}{2}}\kappa,\\
    \hat{r} - \hat{R}&\to\beta\Gamma,
\end{aligned}
\end{equation}
where the new set of order parameters includes $q$, $\chi$, $\hat{Q}$, $\hat{\chi}$, $\tilde{r}$, $\xi$, $\kappa$, and $\Gamma$, which are all of order one in magnitude. In the above scaling, one can derive a finite value of free energy density, from which the saddle point equation can also be derived (details are shown in Appendix~\ref{app-zeroT}).

The qualitative behavior of the new order parameter set is similar to that observed in the large $\beta$ case. The activity order parameter $q$ becomes non-zero when the transition point is crossed to right hand side, while $\chi$ has a sharp decrease when $g$ goes below the transition point. The behavior of $\tilde{r}$ is qualitatively the same with that of $q$. The order parameter $-\xi=\lim_{\beta\to\infty}\sqrt{\beta}(r-R)$ displays a peak at the transition point.

\begin{figure}
    \centering
    \includegraphics[width=1.0\textwidth]{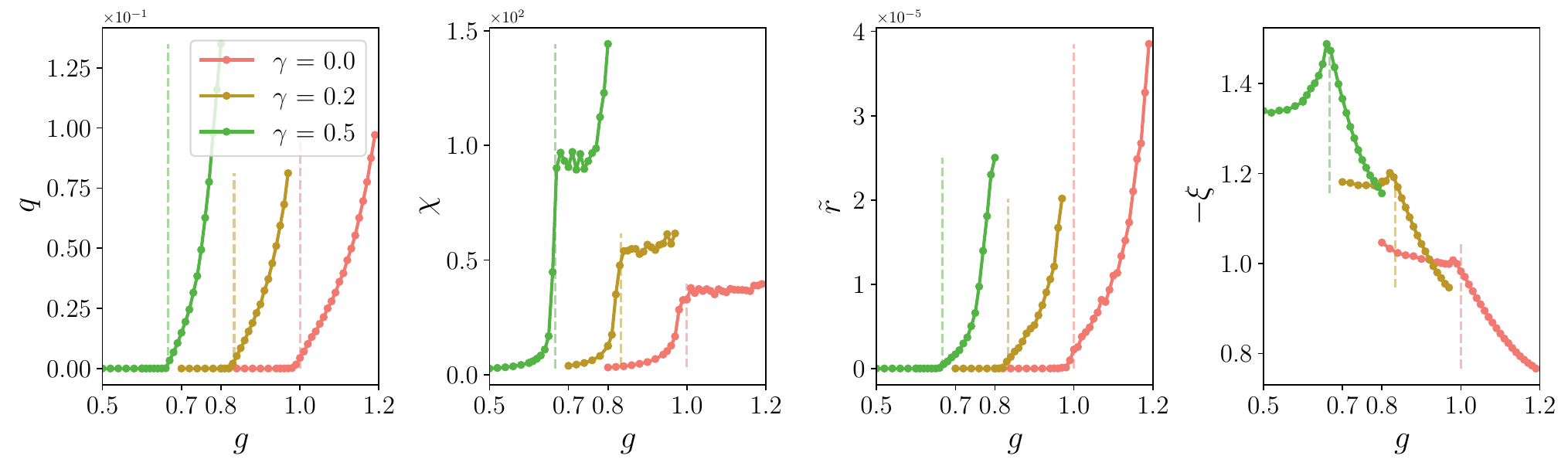}
    \caption{Order parameters as a function of increasing $g$ for different values of $\gamma$. The results are obtained by solving the saddle-point equation in the zero temperature limit [see Eq.~\eqref{sdeT0} in Appendix~\ref{app-zeroT}]. 
   }
    \label{figsdeT0}
\end{figure}

\section{Concluding remarks}
In this paper, we provide a relatively simple way to study non-equilibrium dynamics, by proposing a quasi-potential. Under this potential, we derive the stochastic dynamics to implement the optimization of searching for the low-speed region of the phase space. In the zero temperature limit, only the steady states are considered, and thus our canonical ensemble theory yields a good description of the original non-potential dynamics in the long time limit. More precisely, the low (zero)-speed landscape of the phase space can be analytically studied, which is relevant to our mechanistic understanding of recurrent neural dynamics, which can be adaptive, hierarchical and noisy.
This equilibrium transformation allows for using the Boltzmann measure to describe the low-speed landscape, as could be derived from the Fokker-Planck equation of the corresponding stochastic dynamics of the optimization~\cite{Risken-1996}. Because of the quenched disorder present in our model, the replica method can be readily adapted as well. 

The new finding is that two physically relevant order parameters are derived from our canonical ensemble calculation. One is activity related, which exactly reproduces the order-to-chaos transition phenomenon, previously derived by using more complicated DMFT and random matrix theory~\cite{Chaos-1988,Stein-1988,Brunel-2018,Zou-2023}. The other is response related, which shows a peak at the very transition point, demonstrating the computation benefits of the edge of chaos. The continuous nature of this chaos transition is confirmed by our theory as well. This transition is related to supersymmetry breaking in a field-theoretic representation of the dynamics in an action~\cite{Parisi-1982}.
Remarkably, we relate the replica overlap matrix to measurable fluctuations and responses in non-equilibrium steady states of non-gradient neural dynamics. 

Our work reveals a quasi-potential landscape behind non-gradient Langevin dynamics (or deterministic dynamics). This could inspire several important future directions. First, a complete picture of this landscape can be studied, as powerful equilibrium approaches can be leveraged, such as large deviation analysis~\cite{LDP-2009}, and landscape geometry analysis~\cite{PRE-2014}. Second, one can ask what is the quasi-potential of a coupled-dynamics system where the dynamics of constituent elements can be divided into several levels of different time scales. 
An enlarged state space is required to introduce. Lastly, one could modify the quasi-potential dynamics to approximate the original non-gradient dynamics with increasing precision.

 To sum up, our quasi-potential method provides a simple yet powerful tool to understand the non-equilibrium steady states, especially those without equilibrium limit (non-gradient stochastic dynamics). Our method  allows for order parameters to be defined in canonical ensemble, determination of the transition type, and characterization of fluctuation-response relationship around the dynamics transition, thereby laying down a mathematical foundation of steady state landscape of high dimensional neural dynamics. We build a solid bridge from equilibrium ensemble theory to non-equilibrium non-gradient dynamics, and this optimization angle as a fresh route towards understanding non-equilibrium brain dynamics would prove fruitful in future studies 
 
 Our approach does not rely on the special form of the non-linear dynamics, as long as the speed of the dynamics can be identified, which already includes a broad range of neural dynamics models~\cite{ND-2014,Huang-2022}. Even if the dynamics contains slowly evolving couplings, we can also treat this as a quenched disorder that can be averaged resulting in a self-averaged free energy, from which transitions can be detected and analyzed. As some types of the dynamics can not be accessed by DMFT, or the resultant integro-differential equations for correlation and response functions are hard to solve, our quasi-potential method is a promising starting point to understand the steady state landscape of general non-equilibrium dynamics~\cite{Huang-2023}, which includes competition dynamics of many species 
in ecosystems~\cite{RMP-1971,May-1972,Fyodorov-2016}, high-dimensional dynamics of random optimization~\cite{Fyodorov-2022,Vivo-2024}, high-dimensional gradient flow in machine learning~\cite{Max-2011}, and adaptive, hierarchical and noisy neural dynamics in neuroscience~\cite{ND-2014,Helias-2018,Helias-2020,Clark-2023,Jiang-2023}.

\appendix
\section{Details of replica calculation}\label{app-rep}
Starting from Eq.~\eqref{eq: partitionFunction}, we perform the Hubbard–Stratonovich (HS) transformation $e^{-\frac{1}{2} a b^2} = \int D \hat{x} e^{\i \sqrt{a} b\hat{x}}$ for $a > 0$, to linearize the quadratic terms and have
\begin{equation}
\begin{aligned}
    Z^n =& \int d \mathbf{x} D \mathbf{\hat{x}} \exp \bqty{\i \sqrt{\beta} \sum_{ia} \hat{x}_i^a \pqty{-x_i^a + \sum_j J_{ij} \phi(x_j^a)} - \beta \eta\sum_a \norm*{\mathbf{x}^a}^2},
\end{aligned}
\end{equation}
where we define $D \hat{x} \equiv e^{-\half \hat{x}^2} d \hat{x}/ \sqrt{2 \pi}$ as the Gaussian measure, and we call $\hat x$ as a response field~\cite{Zou-2023}. We define $i,j$ as the neuron index and $a,b$ as the replica index. The neuron index runs from $1$ to $N$, while the replica index runs from $1$ to $n$. In addition, $d \mathbf{x}\equiv\prod_{i, a} d x_i^a$ and $D \mathbf{\hat{x}}\equiv\prod_{i, a} D \hat{x}_i^a$.

Before implementation of the quenched disorder average over the coupling statistics, it is convenient to separate the connection matrix into a symmetric part $\mathbf{J}^S$ and an anti-symmetric part $\mathbf{J}^A$~\cite{Haim-1987},
\begin{equation}\label{eq: JsJa}
    \mathbf{J} = \mathbf{J}^S + \rho \mathbf{J}^A,
\end{equation}
where by definition $J^S_{ij} = J^S_{ji}$ and $J^A_{ij} = - J^A_{ji}$, and their variances are defined as follows,
\begin{equation}\label{eq: varianceJsJa}
    \lrbka{\qty(J^S_{ij})^2} = \lrbka{\qty(J^A_{ij})^2} = \frac{g^2}{N} \frac{1}{1 + \rho^2}.
\end{equation}
Both parts are uncorrelated. It then follows that
\begin{equation}
    \lrbka{J_{ij} J_{ji}} = \frac{g^2}{N} \frac{1 - \rho^2}{1 + \rho^2} = \frac{g^2}{N} \gamma,
\end{equation}
where the relationship between $\rho$ and $\gamma$ is given below,
\begin{equation}
    \rho = \sqrt{\frac{1 - \gamma}{1 + \gamma}}.
\end{equation}

Defining $\lrbka{\cdot}$ as the quenched disorder average, we have
\begin{equation}\label{eq:Z^n}
    \lrbka{Z^n} = \int d \mathbf{x} D \mathbf{\hat{x}} \exp \bqty{-\i \sqrt{\beta} \sum_{ia} x_i^a \hat{x}_i^a - \beta \eta\sum_a \norm*{\mathbf{x}^a}^2} \lrbka{\exp \bqty{ \i \sqrt{\beta} \sum_{ia} \hat{x}_i^a \sum_j J_{ij} \phi(x_j^a)}}.
\end{equation}
The quenched disorder average is first completed as follows,
\begin{small}
\begin{equation}
\begin{aligned}
    & \lrbka{\exp \qty[ \i \sqrt{\beta} \sum_{ij} J_{ij} \sum_a \hat{x}_i^a \phi(x_j^a)]} \\
    =& \lrbka{\exp \qty[ \i \sqrt{\beta} \sum_{ij} \qty(J^s_{ij} + \rho J^a_{ij}) \sum_a \hat{x}_i^a \phi(x_j^a)]} \\
    =& \exp \qty[-\frac{1}{2} \beta \frac{g^2}{N} \frac{1}{1 + \rho^2} \sum_{i < j} \sum_{ab} \qty(\hat{x}_i^a \hat{x}_i^b \phi(x_j^a) \phi(x_j^b) + \hat{x}_i^a \hat{x}_j^b \phi(x_j^a) \phi(x_i^b) + \hat{x}_j^a \hat{x}_i^b \phi(x_i^a) \phi(x_j^b) + \hat{x}_j^a \hat{x}_j^b \phi(x_i^a) \phi(x_i^b))] \\
    &\times \exp \qty[-\frac{1}{2} \beta \frac{g^2}{N} \frac{\rho^2}{1 + \rho^2} \sum_{i < j} \sum_{ab} \qty(\hat{x}_i^a \hat{x}_i^b \phi(x_j^a) \phi(x_j^b) - \hat{x}_i^a \hat{x}_j^b \phi(x_j^a) \phi(x_i^b) - \hat{x}_j^a \hat{x}_i^b \phi(x_i^a) \phi(x_j^b) + \hat{x}_j^a \hat{x}_j^b \phi(x_i^a) \phi(x_i^b))] \\
    =& \exp \qty[-\frac{1}{2} \beta \frac{g^2}{N} \sum_{i \not= j} \sum_{ab} \qty(\frac{1}{1 + \rho^2} \qty(\hat{x}_i^a \hat{x}_i^b \phi(x_j^a) \phi(x_j^b) + \hat{x}_i^a \hat{x}_j^b \phi(x_j^a) \phi(x_i^b)) + \frac{\rho^2}{1 + \rho^2} \qty(\hat{x}_i^a \hat{x}_i^b \phi(x_j^a) \phi(x_j^b) - \hat{x}_i^a \hat{x}_j^b \phi(x_j^a) \phi(x_i^b)))] \\
    =& \exp \qty[-\frac{1}{2} \beta \frac{g^2}{N} \sum_{i j} \sum_{ab} \qty(\hat{x}_i^a \hat{x}_i^b \phi(x_j^a) \phi(x_j^b) + \gamma \hat{x}_i^a \hat{x}_j^b \phi(x_j^a) \phi(x_i^b))] \\
    =& \exp \qty[-\frac{1}{2} \beta g^2 \sum_{ab} Q^{ab} \sum_{i} \hat{x}_i^a \hat{x}_i^b - \frac{1}{2} \beta g^2 N \gamma \sum_{ab} R^{ab} R^{ba}],
\end{aligned}
\end{equation}
\end{small}
where we are left with only $\gamma$ (whose value can now take from $-1$ to $1$, although $\gamma=-1$ is ill-defined for $\rho$), and we treat in our model
the diagonal elements of the connection matrix negligible compared to the contribution from off-diagonal elements of the connection matrix, because of large $N$ limit. To arrive at the last step, we have to introduce the following order parameters
\begin{align}
    Q^{ab} =& \frac{1}{N} \sum_i \phi(x_i^a) \phi(x_i^b), \\
    R^{ab} =& \frac{1}{N} \sum_i \hat{x}_i^a \phi(x_i^b).
\end{align}

Hence, we need to insert the following identity in Eq.~\eqref{eq:Z^n},
\begin{equation}
\begin{aligned}
    1 =& \prod_{a \le b} \int d Q^{ab} \delta \qty(Q^{ab} - \frac{1}{N} \sum_i \phi(x_i^a) \phi(x_i^b)) \prod_{a b} \int dR^{ab} \delta \qty(R^{ab} - \frac{1}{N} \sum_i \hat{x}_i^a \phi(x_i^b)) \\
    =& \int \frac{d \mathbf{Q} d \mathbf{\hat Q} d \mathbf{R} d \mathbf{\hat R}}{(2 \pi)^D} \exp \qty[-\i \sum_{a \le b} Q^{ab} \hat Q^{ab} -\i \sum_{ab} R^{ab} \hat{R}^{ab} + \i \frac{1}{N} \sum_i \sum_{a \le b} \hat Q^{ab} \phi(x_i^a) \phi(x_i^b) + \i \frac{1}{N} \sum_{ab} \hat{R}^{ab} \sum_i \hat{x}_i^a \phi(x_i^b)] \\
    =& \int \frac{d \mathbf{Q} d \mathbf{\hat Q} d \mathbf{R} d \mathbf{\hat R}}{[2 \pi (\i / N)^2]^D} \exp \qty[ - N \sum_{a \le b} Q^{ab} \hat Q^{ab} -N \sum_{ab} R^{ab} \hat{R}^{ab}+ \sum_i \sum_{a \le b} \hat Q^{ab} \phi(x_i^a) \phi(x_i^b)  + \i \sum_{ab} \hat{R}^{ab} \sum_i \hat{x}_i^a \phi(x_i^b)],
\end{aligned}
\end{equation}
where $D=n^2+n(n+1)/2$, and we apply the Fourier transformation of Dirac delta function by introducing the conjugated order parameters $\hat{Q}^{a b}$, $\hat{R}^{a b}$, and rescale the order parameters as $\hat Q^{ab} \to -\i N \hat Q^{ab}$, $R^{ab} \to -\i R^{ab}$ and $\hat{R}^{ab} \to N \hat{R}^{ab}$ in the last step. $d \mathbf{Q} d \mathbf{\hat Q} d \mathbf{R} d \mathbf{\hat R}$ is a shorthand for $\prod_{a \le b} d Q^{ab} d \hat{Q}_{a b} \prod_{a b} d R^{ab} d \hat{R}^{a b}$. 

As a result, Eq.~\eqref{eq:Z^n} becomes
\begin{equation}\label{eq: Z^n2}
\begin{aligned}
    \lrbka{Z^n} \propto& \int d \mathbf{x} D \mathbf{\hat{x}} d \mathbf{Q} d \mathbf{\hat Q} d \mathbf{R} d \mathbf{\hat R}\exp \left[-\i \sqrt{\beta} \sum_{ia} x_i^a \hat{x}_i^a - \beta \eta \sum_{ia} \pqty{x_i^a}^2 -\half g^2 \beta \sum_{ab} Q^{ab} \pqty{\sum_i \hat{x}_i^a \hat{x}_i^b} +\frac{1}{2} \beta g^2 N \gamma \sum_{ab} R^{ab} R^{ba} \right. \\
    & \left. - N \sum_{a \le b} Q^{ab} \hat{Q}^{ab} -N \sum_{ab} R^{ab} \hat{R}^{ab} + \sum_i \sum_{a \le b} \hat{Q}^{ab} \phi(x_i^a) \phi(x_i^b) + \i \sum_{ab} \hat{R}^{ab} \sum_i \hat{x}_i^a \phi(x_i^b)\right], 
\end{aligned}
\end{equation}
where we have neglected the irrelevant constant due to the variable change in the integral, which does not affect the later saddle point approximation
in the large $N$ limit.

Equation~\eqref{eq: Z^n2} can be rewritten into a compact form,
\begin{equation}
    \lrbka{Z^n} \propto \int d \mathbf{Q} d \mathbf{\hat Q} d \mathbf{R} d \mathbf{\hat R} \exp \bqty{N \qty(- \sum_{a \le b} Q^{ab} \hat{Q}^{ab} -\sum_{ab} R^{ab} \hat{R}^{ab} + \G)},
\end{equation}
where the model-dependent action $\G$ reads,
\begin{equation}
\begin{aligned}
    \G =& \ln \int d \mathbf{x} D \mathbf{\hat{x}} \exp \left[-\i \sqrt{\beta} \sum_{a} x^a \hat{x}^a - \beta \eta \sum_{a} \pqty{x^a}^2 - \half g^2 \beta \sum_{ab} Q^{ab}  \hat{x}^a \hat{x}^b + \sum_{a \le b} \hat{Q}^{ab} \phi(x^a) \phi(x^b)\right. \\
    &\left. +\frac{1}{2} \beta g^2 \gamma \sum_{ab} R^{ab} R^{ba} + \i \sum_{ab} \hat{R}^{ab}  \hat{x}^a \phi(x^b)\right],
\end{aligned}
\end{equation}
where $d \mathbf{x}$ and $D \mathbf{\hat{x}}$ have been reduced to $\prod_{a} d x^a$ and $\prod_{a} D \hat{x}^a$, respectively.

To proceed,  we have to assume the replica symmetric (RS) ans\"atz as the first level of approximation, i.e., the replica overlap matrix is permutation invariant, as is intuitively expected from the replicated partition function. This first level of approximation can be refined by going to multiple steps of replica symmetry breaking~\cite{Mezard-1987}, provided that the theoretical prediction is in significant disagreement with numerical simulations. The RS ans\"atz reads  $Q^{ab} = q \delta_{ab} + Q(1 - \delta_{ab})$, $\hat{Q}^{ab} = \hat q \delta_{ab} + \hat{Q}(1 - \delta_{ab})$, $R^{ab} = r \delta_{ab} + R(1 - \delta_{ab})$ and $\hat{R}^{ab} = \hat{r} \delta_{ab} + \hat{R}(1 - \delta_{ab})$. We have then the following result of the averaged replicated partition function,
\begin{equation}
    \lrbka{Z^n} \propto \int d Q d \hat{Q} dq d\hat q d R d \hat{R} dr d\hat{r} \exp \bqty{N \pqty{- \frac{n(n-1)}{2} Q \hat{Q} -n q \hat{q} -n (n-1) R \hat{R} - n r \hat{r} + \G}},
\end{equation}
where
\begin{equation}\label{eq: G}
\begin{aligned}
    \G =& \ln \int d \mathbf{x} D \mathbf{\hat{x}} \exp \left[ -\i \sqrt{\beta} \sum_a x^a \hat{x}^a - \beta \eta \sum_a (x^a)^2 +\frac{1}{2} \beta g^2 \gamma \qty(n(n-1) R^2 + n r^2) - \half g^2 \beta \pqty{Q \pqty{\sum_a \hat{x}^a}^2 + \pqty{q-Q} \sum_a \pqty{\hat{x}^a}^2} \right.\\
    & \left. + \half \pqty{\hat{Q} \pqty{\sum_a \phi(x^a)}^2 + \pqty{2\hat{q} - \hat{Q}} \sum_a \phi^2(x^a)} +\i \qty(\hat{R} \sum_{a b} \hat{x}^a \phi(x^b) + \qty(\hat{r} - \hat{R})\sum_a\hat{x}^a \phi(x^a)) \right].
\end{aligned}
\end{equation}

To linearize the quadratic terms in Eq.~\eqref{eq: G}, we use the HS transformation once again as follows,
\begin{equation}
    e^{\half \hat{Q} \pqty{\sum_a \phi(x^a)}^2 - \half g^2 \beta Q \pqty{\sum_a \hat{x}^a}^2 + \i \hat{R} \sum_{a b} \hat{x}^a \phi(x^b)} = \int Du Dv e^{\kappa_1 u \sum_a \phi(x^a) + v \qty(\kappa_2 \sum_a \phi(x^a) + \i \kappa_3 \sum_a \hat{x}^a)},
\end{equation}
where the auxiliary coefficients $\kappa_1 = \sqrt{\frac{g^2 \beta Q \hat{Q} - \hat{R}^2}{g^2 \beta Q}}$, $\kappa_2 = \frac{\hat{R}}{g \sqrt{\beta Q}}$ and $\kappa_3 = g \sqrt{\beta Q}$. 

Therefore, we have
\begin{equation}
    \G = \ln \int D u D v\exp \left[ \frac{1}{2} \beta g^2 \gamma \qty(n(n-1) R^2 + n r^2)\right]I^n,
\end{equation}
where 
\begin{equation}
\begin{aligned}
I&\equiv\int d x D \hat{x}\exp\left[-\beta\eta x^2-\i\sqrt{\beta}x\hat{x} - \frac{1}{2} g^2 \beta \pqty{q-Q}  \hat{x}^2 + \frac{1}{2} \pqty{2\hat{q} - \hat{Q}}  \phi^2(x) +\i \qty(\hat{r} - \hat{R}) \hat{x} \phi(x)\right.\\
& \left. +  \i \kappa_3 v  \hat{x} + \qty(\kappa_1 u + \kappa_2 v) \phi(x)\right].
\end{aligned}
\end{equation}
 We next complete the integral over $D \hat{x}$ as
\begin{equation}
    \int D \hat{x} \exp \bqty{- \frac{1}{2} g^2 \beta \pqty{q-Q} \hat{x}^2 + \i \qty(\kappa_3 v + \qty(\hat{r} - \hat{R}) - \sqrt{\beta} x) \hat{x}} = \frac{1}{\sigma} \exp \bqty{- \half \frac{1}{\sigma^2} \qty(\kappa_3 v + (\hat{r} - \hat{R}) \phi(x) - \sqrt{\beta} x)^2},
\end{equation}
where $\sigma \equiv \sqrt{1 + g^2 \beta (q-Q)}$. Thus we obtain,
\begin{equation}
    \G = -n \ln \sigma + \frac{1}{2} \beta g^2 \gamma \qty(n (n-1) R^2 + nr^2) + \ln \int D u D v I^n,
\end{equation}
where $I \equiv \int d x e^{\mathcal{H} (x)}$, and
\begin{equation}
    \mathcal{H} (x) \equiv - \beta \eta x^2  + \frac{1}{2} \pqty{2\hat{q} - \hat{Q}} \phi^2(x) + \qty(\kappa_1 u + \kappa_2 v) \phi(x) - \half \frac{1}{\sigma^2} \qty(\kappa_3 v + (\hat{r} - \hat{R}) \phi(x) - \sqrt{\beta} x)^2,
\end{equation}
where $\mathcal{H}(x)$ could be considered as the single-variable effective Hamiltonian of our model, which emerges after the quenched disorder is averaged out. 

Finally, we perform the saddle point approximation because of large $N$, such that $\lrbka{Z^n} \propto \int d \mathcal{O} d \hat{\mathcal{O}} e^{N \Phi(\mathcal{O}, \hat{\mathcal{O}})} \approx e^{N \Phi(\mathcal{O}^* ,\hat{\mathcal{O}}^*)}$, where $\mathcal{O} = \{Q , q, R, r\}$ and the superscript $*$ refers to the saddle point value. The free energy can be obtained from the replica trick $- \beta f = \lim_{n \to 0} \frac{\ln \lrbka{Z^n}}{nN}$, and therefore,
\begin{equation}\label{eq: free energy equation}
    - \beta f = \frac{1}{2} Q \hat{Q} - q \hat{q} +  R \hat{R} - r \hat{r} - \ln \sigma + \frac{1}{2} \beta g^2 \gamma \qty(r^2 - R^2) + \int D u D v \ln I.
\end{equation}

\section{Saddle Point Equations}\label{app-sde}
In this section, we omit the superscript $*$ on the order parameters, and derive the self-consistent equations these order parameters must obey, which is the so-called saddle point equations (SDEs). The principle is to optimize the action with respect to the order parameters and their conjugated counterparts. The result is given below, 
\begin{subequations} \label{eq: SDEs}
\begin{align}
    q =& \qty[\lrbka{\phi^2}] ,\\
    Q =& \qty[\lrbka{\phi}^2], \\
    r =& -\frac{1}{\sigma^2} f(1, 0, -1) \qty[\lrbka{\phi^2}] + \frac{1}{\sigma^2} f(0, 1, -1) \qty[\lrbka{\phi}^2] + \frac{\sqrt{\beta}}{\sigma^2} \qty(1 - g k Q) \qty[\lrbka{x \phi}] + \frac{\sqrt{\beta}}{\sigma^2} g k Q \qty[\lrbka{x}\lrbka{\phi}], \\
    R =& -\frac{1}{\sigma^2} f(0, 1, -1) \qty[\lrbka{\phi^2}] - \frac{1}{\sigma^2} f(1, -2, 1) \qty[\lrbka{\phi}^2] - \frac{\sqrt{\beta}}{\sigma^2} g k Q \qty[\lrbka{x \phi}] + \frac{\sqrt{\beta}}{\sigma^2} \qty(1 + g k Q) \qty[\lrbka{x}\lrbka{\phi}],\\
    \hat q =& -\frac{g k}{2} + \frac{g^2 k^2 Q}{2} + \frac{g k}{2 \sigma^2} \qty(\hat{r} -\hat{R}) f(1, 1, -2) \qty[\lrbka{\phi^2}] + \frac{g k}{\sigma^2} \qty(\hat{r} -\hat{R}) f(0, -1, 1) \qty[\lrbka{\phi}^2] + \frac{k^2}{2} \qty(1 - 2 g k Q) \qty[\lrbka{x^2}] \\
    + &\frac{g k \sqrt{\beta}}{\sigma^2} f (0, 1, -2) \qty[\lrbka{x}\lrbka{\phi}] + \frac{g k \sqrt{\beta}}{\sigma^2} f(-1, 0, 2) \qty[\lrbka{x \phi}] + g k^3 Q \qty[\lrbka{x}^2], \\
    \hat Q =& g^2 k^2 Q + \frac{2 g k}{\sigma^2} \qty(\hat{r} -\hat{R}) f(0, 1, -1) \qty[\lrbka{\phi^2}] +\frac{g k}{\sigma^2} \qty(\hat{r} -\hat{R}) f(1, -3, 2) \qty[\lrbka{\phi}^2] - 2 g k^3 Q \qty[\lrbka{x^2}] \\
    - &\frac{2 g k \sqrt{\beta}}{\sigma^2} f(1, -2, 2) \qty[\lrbka{x}\lrbka{\phi}] + \frac{2 g k \sqrt{\beta}}{\sigma^2} f(0, -1, 2) \qty[\lrbka{x \phi}] + k^2 \qty(1 + 2 g k Q) \qty[\lrbka{x}^2] ,\\
    \hat{r} =& \beta g^2 \gamma r, \\
    \hat{R} =& \beta g^2 \gamma R,
\end{align}
\end{subequations}
where $k \equiv \frac{g \beta}{\sigma^2}$ and $f (a, b, c) \equiv a \hat{r} + b \hat{R} + c g k Q \qty(\hat{r} - \hat{R})$, and we define disorder and thermal averages, respectively as follows,
\begin{equation}
    \qty[\bullet] \equiv \int Du Dv \bullet,
\end{equation}
and
\begin{equation}\label{eq: internalIntegration}
    \lrbka{ \bullet } \equiv \frac{\int dx e^{\mathcal{H}(x)} \bullet}{\int dx e^{\mathcal{H}(x)}}.
\end{equation}
 Note that we have used the integral identity $\int D x x f(x) = \int D x f'(x)$ where $f(x)$ is a bounded and differentiable function to write the above SDEs in the compact form.
 
 We are also interested in the average energy, whose ground state value is related to the speed level of the original dynamics. The thermodynamic relation $\lrbka{E}=\frac{\partial[\beta f]}{\partial\beta}$ is used to derive the following formula,
 \begin{equation}
\begin{aligned}
    U =& \frac{g^2}{2 \sigma^2} \qty(q - g k Q (q- Q)) - \half g^2 (r^2 - R^2) \gamma +\frac{1}{2 \sigma^2} \qty(1 + 2 \eta \sigma^2 - g k (q - Q) - \frac{2 g k Q}{\sigma^2}) \qty[\lrbka{x^2}] + \frac{g k Q}{\sigma^4} \qty[\lrbka{x}^2] \\
    & - \frac{1}{2 \sqrt{\beta} \sigma^2} \qty(f(1, -1, 0) + \frac{1}{\sigma^2} f(0, 1, -3) + g k (q - Q) f(-2, 1, 1)) \qty[\lrbka{x \phi}] \\
    & - \frac{1}{2 \sqrt{\beta} \sigma^2} \qty(\frac{1}{\sigma^2} f(0, -1, 3) + g k (q - Q) f(0, 1, -1)) \qty[\lrbka{x} \lrbka{\phi}] \\
    & - \frac{1}{2 g^2 Q \beta^2} \qty(\hat{R}^2 + g^2 k^2 Q (q - Q) \qty(\hat{r} - \hat{R})^2 + f(0, -1, 1) \qty(f(0, 1, 1) - 2 g^2 k^2 Q (q - Q) (\hat{r} - \hat{R}))) \qty[\lrbka{\phi^2}] \\
    & + \frac{1}{2 g^2 Q \beta^2} \qty(\hat{R}^2
    + f(0, -1, 1) \qty(f(0, 1, 1) - 2 g^2 k^2 Q (q - Q) (\hat{r} - \hat{R}))) \qty[\lrbka{\phi}^2].
\end{aligned}
\end{equation}
To characterize the average activity level in the neural population, we also derive the $\ell_2$ norm of the activity as follows
\begin{equation}
     \frac{1}{N}\lrbka{\|\mathbf{x}\|^2} = - \frac{1}{\beta} \pdv{\qty(- \beta f)}{\eta} = \qty[\lrbka{x^2}].
\end{equation}

\section{Independent case $\gamma = 0$}\label{app-0}
In the case $\gamma = 0$, all the connection weights are independent and uncorrelated. Hence, the expression of the SDEs gets greatly simplified as
\begin{align} \label{eq: SDEsGamma0.0}
    q =& \qty[\lrbka{\phi^2}] ,\\
    Q =& \qty[\lrbka{\phi}^2] ,\\
    r =& \frac{\sqrt{\beta}}{\sigma^2} \qty(1 - g k Q) \qty[\lrbka{x \phi}] + \frac{\sqrt{\beta}}{\sigma^2} g k Q \qty[\lrbka{x}\lrbka{\phi}], \\
    R =& - \frac{\sqrt{\beta}}{\sigma^2} g k Q \qty[\lrbka{x \phi}] + \frac{\sqrt{\beta}}{\sigma^2} \qty(1 + g k Q) \qty[\lrbka{x}\lrbka{\phi}] ,\\
    \hat q =& -\frac{g k}{2} + \frac{g^2 k^2 Q}{2} + \frac{k^2}{2} \qty(1 - 2 g k Q) \qty[\lrbka{x^2}] + g k^3 Q \qty[\lrbka{x}^2] ,\\
    \hat Q =& g^2 k^2 Q - 2 g k^3 Q \qty[\lrbka{x^2}] + k^2 \qty(1 + 2 g k Q) \qty[\lrbka{x}^2], \\
    \hat{r} =& 0, \\
    \hat{R} =& 0.
\end{align}
Because $\hat{r} = \hat{R} = 0$, we have $f (a, b, c) = 0$. The free energy is thus reduced to
\begin{equation}
    - \beta f = \frac{1}{2} Q \hat{Q} - q \hat{q} - \ln \sigma - \frac{g^2 \beta}{2 \sigma^2} Q + \int D u D v \ln I,
\end{equation}
where
\begin{equation}
    I \equiv \int d x \exp \bqty{- \half \beta \pqty{\frac{1}{\sigma^2} + 2 \eta} x^2 + \frac{1}{2} \pqty{2\hat q - \hat Q} \phi^2(x)  + \sqrt{\hat Q} u \phi(x) +  \frac{g \beta}{\sigma^2} \sqrt{Q} v x}.
\end{equation}
 This free energy does not depend on the order parameters $r$ and $R$ anymore.
The average energy also yields the following simple form
\begin{equation}
    U = \frac{g^2}{2 \sigma^2} \qty(q - g k Q (q- Q)) +\frac{1}{2 \sigma^2} \qty(1 + 2 \eta \sigma^2 - g k (q - Q) - \frac{2 g k Q}{\sigma^2}) \qty[\lrbka{x^2}] + \frac{g k Q}{\sigma^4} \qty[\lrbka{x}^2].
\end{equation}

\section{Additional experiment results}\label{app-exp}
\begin{figure*}[ht]
    \centering
    \includegraphics[width=1.0\textwidth]{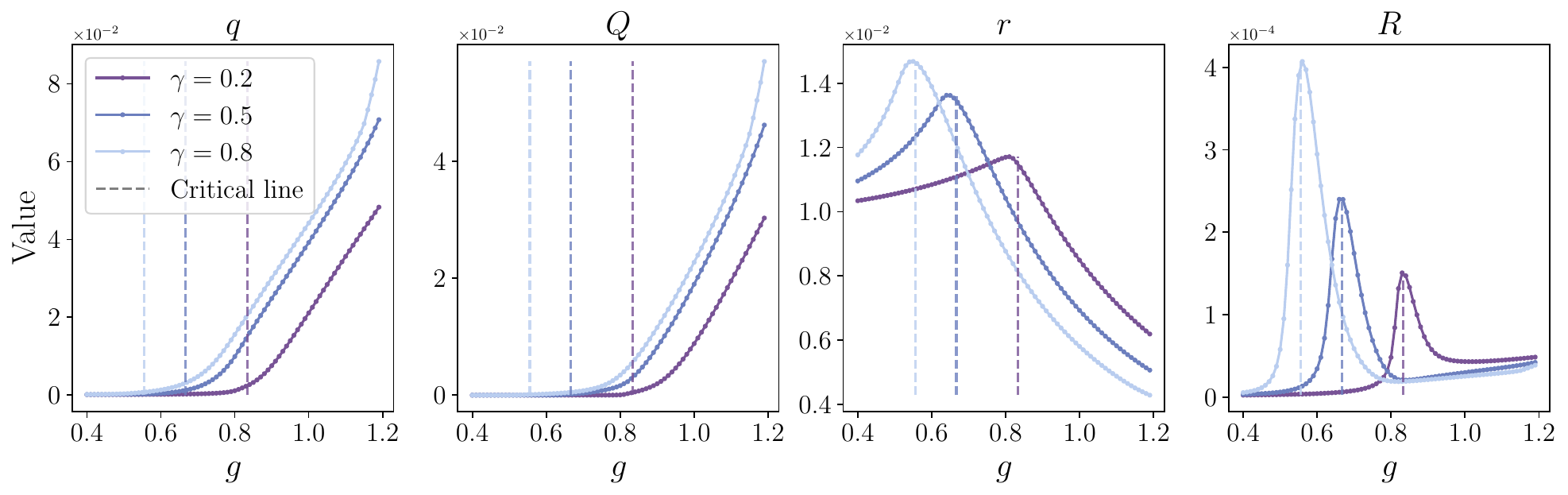}
    \caption{Phase transitions as $g$ increases. 
    We plot three different curves ($\gamma = 0.2$, $0.5$, $0.8$), and different colors represent different $\gamma$ values. 
    The dashed line represents the DMFT result of critical points in Ref.~\cite{Brunel-2018}.}
    \label{PD-2d}
\end{figure*}

To illustrate the phase transition in more details, we choose some curves ($\gamma = 0.2$, $0.5$, $0.8$) and plot them in the 2-dimension plane (Fig.~\ref{PD-2d}). $q$ and $Q$ are close to zero before the critical point, while growing rapidly after the critical point. $r$ and $R$ achieve their maximum values at the critical point. This is also supported by the iteration dynamics of the SDEs (Fig.~\ref{iter}).

\begin{figure}[ht]
    \centering
    \includegraphics[width=0.4\textwidth]{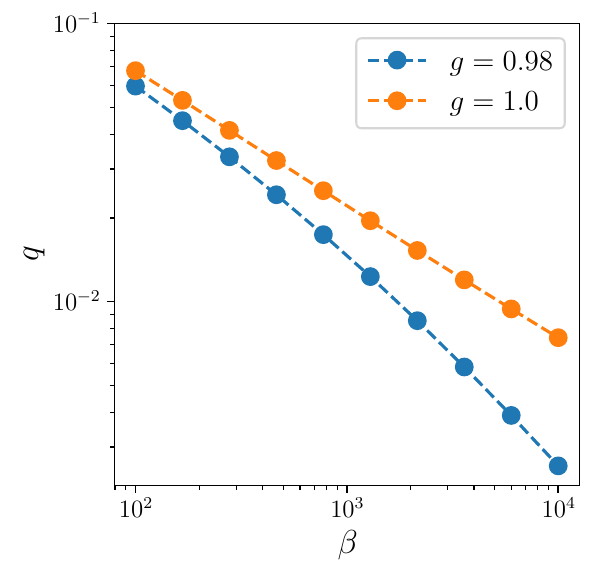}
    \caption{The relationship between $q$ and $\beta$ when $g = 0.98$ and $g = 1.0$ ($\gamma=0$). The blue dots denote $g = 0.98$ case, and the orange dots denote $g = 1.0$. Both axes are in the log scale.}
    \label{qbeta}
\end{figure}

\begin{figure}[ht]
    \centering
    \includegraphics[width=0.8\textwidth]{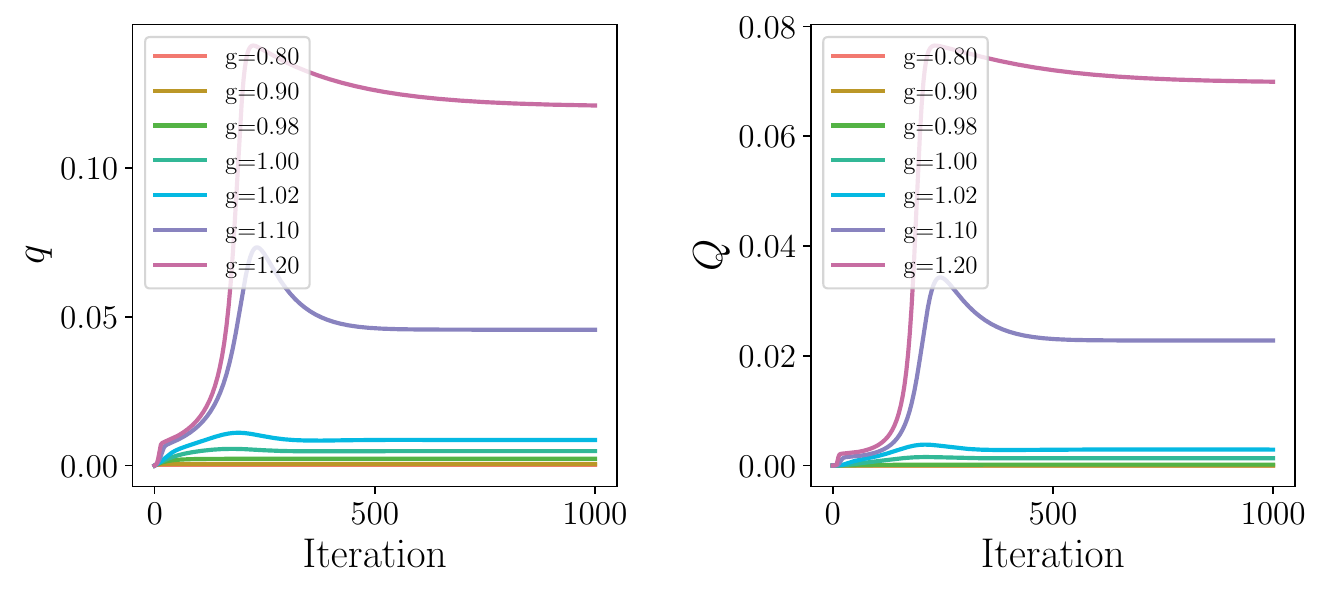}
    \caption{The iterative dynamics of $q$ and $Q$ for different values of $g$. All initial values are set to $10^{-5}$ for $q$ and $10^{-6}$ for $Q$.
    The data points are obtained by using the Algo.~\ref{alg}.}
    \label{iter}
\end{figure}

Next we explore the effect of $\beta$ in optimizing the energy and the order parameters. When $g$ is close to the critical point, we show how $q$ varies with increasing $\beta$ (see Fig.~\ref{qbeta} for $g = 0.98$ and $g = 1.0$). We observe that $q$ converges more rapidly towards zero when $g$ is below the critical value. When $g$ is approaching the critical value from below, the decaying towards zero becomes much slower, signaling a continuous phase transition.

\section{Numerical details of solving SDEs}\label{app-sol}
The most expensive cost of solving SDEs comes from 
the  double average terms, such as $\qty[\lrbka{x}^2]$ and $\qty[\lrbka{\phi}^2]$, where the outer integral is related to $D u D v$ and the inner integral is related $d x$. 
These integrals are calculated by Monte Carlo (MC) method with the important sampling technique~\cite{Huang-2022}. 
Therefore, we need to generate $100k\times6k$ Gaussian samples, where $100k$ MC samples of $u$ and $v$ is used for the outer integral and $6k$ MC samples of $x$ for each pair of $u$ and $v$, is used for the inner integral. $k$ here represents $1\,000$.
Note that the inner integral in Eq.~\eqref{eq: internalIntegration} does not contain Gaussian measure at first sight. However, we can take a reorganization of terms in the effective Hamiltonian.  We have then the following equivalent transformation,
\begin{equation}\label{eq: internalIntegration2}
    \lrbka{ \bullet } \equiv \frac{\int Dx e^{\tilde{\mathcal{H}}(x)} \bullet}{\int Dx e^{\tilde{\mathcal{H}}(x)}},
\end{equation}
where the Gaussian measure is introduced as $D x \equiv e^{-\half \beta \qty(\frac{1}{\sigma^2} + 2 \eta) x^2} d x/ \sqrt{2 \pi }$ and
\begin{equation}
    \tilde{\mathcal{H}} (x) \equiv \frac{1}{2} \pqty{2\hat{q} - \hat{Q}} \phi^2(x) + \qty(\kappa_1 u + \kappa_2 v) \phi(x) - \half \frac{1}{\sigma^2} \qty(\kappa_3 v + (\hat{r} - \hat{R}) \phi(x))^2 + \frac{\sqrt{\beta}}{\sigma^2} \qty(\kappa_3 v + (\hat{r} - \hat{R}) \phi(x)) x .
\end{equation}

After initialization of the order parameters, we complete the double average using the above MC sampling. Then we update the order parameters as follows,
\begin{equation}
    \mathcal{O}_{t+1} = \alpha \mathcal{O}_t + (1 - \alpha) f(\mathcal{O}_t),
\end{equation}
where $\alpha$ is the damping parameter (a value close to one) to speed up the convergence. $f(\mathcal{O})$ represents the right-hand-side of Eqs.~\eqref{eq: SDEs}. We stop the iteration when $\abs{\mathcal{O}_{t+1} - \mathcal{O}_t} < 10^{-3}$ for all order parameters.
We give a pseudocode in Algorithm.~\ref{alg} for solving the SDEs.

\begin{algorithm}[h]
\caption{Solver of SDEs}
\begin{algorithmic}[1]
\Require The initial values of order parameters $\mathcal{O}_0$, $\mathcal{O} \in \{q, Q, r, R\}$
\Ensure The convergent values of order parameters $\mathcal{O}^*$
\State Generate $100k \times 6k$ Gaussian Monte Carlo samples
\Repeat
    \State Calculate all the double averages by the MC method in Eqs.~\eqref{eq: SDEs}
    \State Let $\mathcal{O}_{t+1} \gets \alpha \mathcal{O}_t + (1 - \alpha) f(\mathcal{O}_t)$, where $f(\mathcal{O})$ are the functions specified by the SDEs [see Eqs.~\eqref{eq: SDEs}]
\Until $\abs{\mathcal{O}_{t+1} - \mathcal{O}_t} < 10^{-3}$
\State Let $\mathcal{O}^* \gets \mathcal{O}_t$
\end{algorithmic}
\label{alg}
\end{algorithm}

\section{Zero temperature analysis}\label{app-zeroT}
Setting $\beta\to\infty$, we focus on the ground state of the quasi-potential, which describes the non-equilibrium steady states of Eq.~\eqref{eq: dynamicEquation}. In the zero temperature limit, we have to consider the following scaling behavior guaranteeing a finite value of free energy when $\beta\to\infty$ [see Eq.~\eqref{fT}].
\begin{equation}
\begin{aligned}
    q - Q&\to\frac{\chi}{\beta}, \\
    \hat{Q} &\to \beta^2 \hat{Q}, \\
    2 \hat{q} - \hat{Q}&\to-2\beta \hat{\chi},\\
    r&\to\sqrt{\beta}\tilde{r},\\
    R-r&\to\frac{\xi}{\sqrt{\beta}},\\
    \hat{R}&\to\beta^{\frac{3}{2}}\kappa,\\
    \hat{r} - \hat{R}&\to\beta\Gamma,
\end{aligned}
\end{equation}
where the new set of order parameters includes $q$, $\chi$, $\hat{Q}$, $\hat{\chi}$, $\tilde{r}$, $\xi$, $\kappa$, and $\Gamma$, which are all of order one in magnitude. 

Inserting the above scaling expressions into the free energy [see Eq.~\eqref{fT}], which then reads
\begin{equation}
\begin{aligned}
    - \beta f =& -\half \beta \qty(-2 q \hat{\chi} + \hat{Q} \chi) - \beta \qty(\tilde{r} \Gamma - \kappa \xi) \\
    & - g^2 \beta \gamma \xi \tilde{r} + \int \pqty{D u D v} \ln \int d x e^{\beta \mathcal{H}_0 (x)},
\end{aligned}
\end{equation}
where
\begin{equation}
\begin{aligned}
    \mathcal{H}_0 (x) =& - \eta x^2 - \hat{\chi} \phi^2(x) \\
    &+ \qty(\sqrt{\hat{Q} - \frac{\kappa^2}{g^2 q}} u + \frac{\kappa}{g \sqrt{q}} v) \phi(x) \\
    &- \half \frac{1}{\sigma^2} \qty(g \sqrt{q} v + \Gamma \phi(x) - x)^2.
\end{aligned}
\end{equation}

In the zero temperature limit $\beta \to \infty$, we apply the Laplace method to estimate the following integral
\begin{equation}
    \int D u D v \ln \int d x e^{\beta \mathcal{H}_0 (x)} = \beta \int D u D v \mathcal{H}_0 (x^*),
\end{equation}
where $x^* = \operatorname{argmax}_x \mathcal{H}_0 (x)$. Note that all terms in $- \beta f$ are proportional to $\beta$, and hence the free energy density in the zero temperature limit is given by
\begin{equation}
\begin{aligned}
    - f =& -\half \qty(-2 q \hat{\chi} + \hat{Q} \chi) - \qty(\tilde{r} \Gamma - \kappa \xi) - g^2 \gamma \xi \tilde{r} \\
    &+ \int D u D v\mathcal{H}_0 (x^*).
\end{aligned}
\end{equation}

The new set of order parameters obeys the self-consistent equation making the zero temperature free energy stationary. We thus get the following saddle-point equation:
\begin{equation}\label{sdeT0}
\begin{aligned}
    q =& \qty[\phi^2(x^*)], \\
    \chi =& \frac{g q}{\sqrt{g^2 q^2 \hat{Q} - q \kappa^2}} \qty[u \phi(x^*)], \\
    \hat{Q} =&  \frac{g^2}{\sigma^4} \left(g^2 q + \Gamma^2 \qty[\phi^2(x^*)] + \qty[\qty(x^*)^2] \right.\\
    & - 2 \Gamma \qty[x^* \phi(x^*)] + 2 g \sqrt{q} \Gamma \qty[v \phi(x^*)] - 2 g \sqrt{q} \qty[v x^*] \Big), \\
    \hat{\chi} =& \frac{g^2}{2 \sigma^2} -  \frac{\kappa^2}{2 g q \sqrt{\hat{Q} g^2 q^2 -  q \kappa^2}} \qty[u \phi(x^*)] \\
    & + \frac{1}{2 \sqrt{q}}\qty(\frac{g \Gamma}{\sigma^2} + \frac{\kappa}{g q}) \qty[v \phi(x^*)] - \frac{1}{ \sigma^2} \frac{g}{2 \sqrt{q}} \qty[v x^*], \\
    \tilde{r} =& - \frac{g \sqrt{q}}{\sigma^2} \qty[v \phi(x^*)] - \frac{\Gamma}{\sigma^2} \qty[\phi^2(x^*)] + \frac{1}{\sigma^2} \qty[x^* \phi(x^*)], \\
    \xi =& \frac{\kappa}{g \sqrt{g^2 q^2 \hat{Q} - q \kappa^2}} \qty[u \phi(x^*)] - \frac{1}{g \sqrt{q}} \qty[v \phi(x^*)], \\
    \kappa =& g^2 \gamma \tilde{r}, \\
    \Gamma =& -g^2 \gamma \xi,
\end{aligned}
\end{equation}
where $\left[ \bullet \right]\equiv\int DuDv$ as before, and to estimate this average, we first generate $M = 5 \times10^6$ Monte Carlo samples $\{(u_i, v_i)\}_{i = 1}^M$, for each of them we find the global maximum of $\mathcal{H}_0 (x)$, i.e., $x^*$. All these values corresponding to the maxima [for each pair of $(u,v)$]  are further used to complete the calculation of the order parameter for one round. Codes of this paper are released on the GitHub link~\cite{JB-2024}.

\begin{acknowledgments}
We would like to thank Zijian Jiang, Yang Zhao and Wenkang Du for helpful discussions at earlier stages of this project. This research was supported by the National Natural Science Foundation of China for
Grant Number 12122515 (H.H.), and Guangdong Provincial Key Laboratory of Magnetoelectric Physics and Devices (No. 2022B1212010008), 
and Guangdong Basic and Applied Basic Research Foundation (Grant No. 2023B1515040023).  
\end{acknowledgments}


\end{document}